\documentclass[a4paper]{jpconf}
\usepackage{graphicx}
\usepackage{amssymb}
\usepackage{amsmath}
\begin{document}
\title{Virial tests for post-Newtonian stationary black-hole--disk systems}

\author{Piotr Jaranowski}
\address{Wydzia\l~ Fizyki, Uniwersytet w Bia{\l}ymstoku, Cio{\l}kowskiego 1L, 15--245 Bia{\l}ystok, Poland}
\ead{p.jaranowski@uwb.edu.pl}
\author{Patryk Mach, Edward Malec, Micha\l~Pir\'og}
\address{{Instytut Fizyki im.~Mariana  Smoluchowskiego, Uniwersytet Jagiello\'nski, {\L}ojasiewicza 11, 30-348 Krak\'{o}w, Poland} }
\ead{patryk.mach@uj.edu.pl, malec@th.if.uj.edu.pl, michal.pirog@uj.edu.pl}

\begin{abstract}
We investigated hydrodynamical post-Newtonian models of selfgravitating stationary black-hole--disk systems. The post-Newtonian scheme presented here and also in our recent paper is a continuation of previous, purely Newtonian studies of selfgravitating hydrodynamical disks rotating according to the Keplerian rotation law. The post-Newtonian relativistic corrections are significant even at the 1PN level. The 1PN correction to the angular velocity can be of the order of 10\% of its Newtonian value. It can be expressed as a combination of geometric and hydrodynamical terms. Moreover, in contrast to the Newtonian Poincar\'{e}--Wavre theorem, it depends both on the distance from the rotation axis and the distance from the equatorial plane.

In the technical part of this note we derive virial relations valid up to 1PN order. We show that they are indeed satisfied by our numerical solutions.
\end{abstract}

\section{Introduction}

In a recent paper \cite{JMMP_1PN} we investigated post-Newtonian models of selfgravitating gaseous disks that rotate according to the Keplerian rotation law. The analysis presented there is a continuation of our previous studies, where such disk systems were investigated in Newtonian theory \cite{MMP_0PNa, MMP_0PNb}.

In the Newtonian framework we posed the following problem: Suppose one observes a selfgravitating stationary gaseous disk around a central object (modeled by a point-mass) that rotates according to the Keplerian rotation law, that is with the angular velocity $v_0^\phi = \omega_0 / r^{3/2}$, where $r$ is the distance from the rotation axis, and $\omega_0$ is a constant. For the disk consisting of test particles we have $\omega_0 = \sqrt{GM_\mathrm{c}}$, where $M_\mathrm{c}$ is the mass of the central object, and $G$ is the gravitational constant. What does the observed value of $\omega_0$ correspond to in the case where the mass of the disk is comparable with $M_\mathrm{c}$? Is it still the central mass $M_\mathrm{c}$, the sum of the two masses, or some nontrivial combination of them? It turns out that the selfgravity speeds up the rotation of the disk---it rotates faster than this would follow from the Keplerian formula involving the central mass $M_\mathrm{c}$ only. Moreover, the way in which $\omega_0$ depends on the central mass and the mass of the disk is prescribed by the geometry of the disk. Thus, in principle, it is possible to measure the masses of Keplerian disks, whenever their geometry is known. This procedure was applied to the accretion disk in the AGN of NGC 4258, where the Keplerian rotation curve was measured in the maser emission \cite{MMP_0PNb}.

In \cite{JMMP_1PN} we extended the Newtonian analysis to the first post-Newtonian approximation (1PN). Selfgravitating stationary gaseous disks were investigated before in full relativity (cf.~\cite{Ni1, Ni2}). We decided on the post-Newtonian scheme, because of its conceptual simplicity. In particular, the notion of the Keplerian rotation has a clear meaning in the post-Newtonian scheme.

The main result obtained in the 1PN approximation is that the angular velocity profile is affected in two different ways---some parts of a disk can be speeded up and the others slowed down. Furthermore, the sum of the Newtonian and post-Newtonian components of the angular velocity is not anymore a function of the cylindrical radius only, but in general it depends on radial and vertical coordinates \cite{JMMP_1PN}.

In this paper we sketch briefly the main equations that constitute the 1PN model and then derive virial-type relation that can be used to test the obtained numerical solutions. Suitable virial tests valid in the Newtonian case were presented in \cite{mach_virial} and \cite{MMP_0PNa}. We discuss them here for clarity. The post-Newtonian virial identities given here are new and have not been discussed in \cite{JMMP_1PN}. In the last section of this paper we also show that they are satisfied by our numerical models with the accuracy similar to that of Newtonian solutions.

\section{Description of the model}

Our 1PN black-hole--disk models are constructed assuming the metric of the form
 \begin{eqnarray}
\label{metric}
ds^2 & = & g_{\mu \nu} d x^\mu d x^\nu = \left( -1-2{ \frac {U \left( x,y,z\right) }{{c}^{2}}}-2{\frac { \left( U\left( x,y,z \right)  \right) ^{2}}{{c}^{4}}}\right) (d x^0)^2
- 2{\frac {A_i\left(x,y,z\right) }{{c}^{3}}} dx^i  dx^0
\nonumber \\
& & + \left(1-2{\frac {U \left( x,y,z \right) }{{c}^{2}}}\right)  \left(d x^2 + d y^2 + d z^2\right),
\end{eqnarray}
where we use Cartesian coordinates $x^0=ct, x^1=x, x^2 = y, x^3 = z$, and $c$ is the speed of light.

We write the energy-momentum tensor as
\begin{equation}
T^{\alpha\beta} =\frac{M_{\textrm{c}} c^2}{\sqrt{g}}
\frac{u^\alpha_\textrm{BH}u^\beta_\textrm{BH}}{u^0_\textrm{BH}}
\delta(\mathbf{x}) + \rho (c^2+h)u^\alpha u^\beta + p g^{\alpha\beta},
\end{equation}
where the first component describes the point particle (it is proportional to the Dirac delta distribution and models the central black hole) at rest, located at the origin of the coordinate system; the second one is the energy-momentum tensor of the disk matter. Here $M_{\textrm{c}}$ denotes the mass of the point particle; $g$ is the determinant of the metric $g = -\det(g_{\mu\nu})$. The four-vectors $u_\mathrm{BH}^\alpha$ and $u^\alpha$ denote the four-velocities of the central point-mass and the fluid, respectively. The symbol $\rho$ denotes the baryonic rest-mass density, $h$ is the  specific enthalpy, and $p$ is the pressure. In the following sections we will also work with the three-velocity, defined as $v^i = c u^i/u^0$, $i = 1, 2, 3$.

In the remaining part of the article we use standard cylindrical coordinates ($r, z, \phi$). We consider a stationary, selfgravitating, axially and equatorially symmetric polytropic disk, rotating around a central point mass $M_\textrm{c}$ according to the Keplerian rotational law $ v_0^{\phi} = \omega_0 r^{-3/2}$. We assume that the disk is geometrically bounded by the inner and outer radius $r_\mathrm{{in}}$ and $r_\mathrm{{out}}$, respectively. We introduce the notation according to which any quantity $\xi$ (if it is necessary) is separated into its Newtonian $\xi_0$ and post-Newtonian $\xi_1$ part according to the general pattern $\xi = \xi_0 + \xi_1 / c^2$. Following \cite{Sch} we derive basic equations which, separated into their Newtonian and post-Newtonian parts, read:
\begin{eqnarray}
\Delta U_0 & = & 4 \pi G \left( \rho_0 + M_\mathrm{c} \delta(\mathbf x) \right) \label{a}, \\
h_0 & = & - U_0 + \int d r (v_0^{\phi})^2 r + C_0 \label{b}, \\
\Delta A_\phi  & = & 2\frac{\partial_rA_\phi }{r} -16\pi G r^2\rho_0 v^\phi_0 \label{c}, \\
\Delta U_1 & = & 4 \pi G \left(M_\mathrm{c} U^\mathrm{D}_0(0) \delta (\mathbf x) + \rho_1 + 2p_0 + \rho_0 \left( h_0 - 2 U_0 + 2 r^2 (v_0^\phi)^2 \right) \right) \label{d}, \\
h_1 & = & - U_1 - v_0^{\phi}A_{\phi} + 2h_0 (v_0^{\phi})^2 r^2 - \int d r (v_0^{\phi})^4 r^3 - \frac{3}{2}h_0^2 - 4h_0U_0 - 2U_0^2 - C_1 \label{e},
\end{eqnarray}
where $\Delta$ denotes the flat Laplacian with respect to coordinates $(x^1, x^2, x^3)$, $C_0$ and $C_1$ are integration constants, and $U^\mathrm{D}_0$ is the gravitational potential due to the disk only, i.e., $U_0 = - G M_\mathrm{c}/|\mathbf{x}| + U^\mathrm{D}_0$. They follow from the conservation law, $\nabla_\alpha T^{\alpha\beta} = 0$, the continuity equation $\nabla_\alpha\left(\rho u^\alpha\right)=0$, and Einstein equations
\begin{equation}
R_{\mu \nu} - \frac{R}{2} g_{\mu \nu } = 8\pi \frac{G}{c^4} T_{\mu \nu },
\label{ee1}
\end{equation}
where $R_{\mu \nu}$ is the Ricci tensor and $R$ denotes the Ricci scalar.

The above system of equations is closed by assuming an equation of state. Our numerical solutions are obtained for a polytropic equation of state of the form $p = K \rho^\gamma$, or equivalently $h_0 = K \gamma/(\gamma - 1) \rho_0^{\gamma - 1}$ (for the $0^{th}$ order solution) and $h_1 = (\gamma - 1)h_0 \rho_1/\rho_0$ (for the 1PN part), where $K$ and $\gamma > 1$ are constants.

Equations (\ref{a}--\ref{e}) should be solved with respect to the Newtonian gravitational potential $U_0(r, z)$, the post-Newtonian gravitational potential $U_1(r, z)$, the rotational potential $A_{\phi}(r, z)$ and the Newtonian and post-Newtonian enthalpy $h_0(r, z)$ and $h_1(r, z)$. Any other quantity (the density $\rho(r, z)$, the pressure $p(r, z)$, etc.) can be obtained from these five functions.

Numerical solutions are obtained as follows. We use the classic Self-Consistent Field (SCF) scheme (cf.~\cite{O_and_M}) to solve the set of Eqs.~(\ref{a}) and Eq.~(\ref{b}). Given the Newtonian potential $Y_0$ and the enthalpy $h_0$, we can solve Eq.~(\ref{c}) at once. Finally, we use again the SCF scheme to solve the set of Eqs.~(\ref{d}) and (\ref{e}). For the Keplerian rotation law the above method converges for all values of the parameters that we have tested.

\section{Results}

The main result of our post-Newtonian scheme is the correction to the Newtonian angular velocity profile. The well-known theorem by Poincar\'{e} and Wavre states that Newtonian stationary barotropic disks (or stars) rotate with the angular velocity that can depend on the distance $r$ from the rotation axis only \cite{tassoul}. It turns out that already the 1PN correction can be significant and in general it depends also on $z$. It can be expressed as 
\begin{equation}
\label{ang_vev_pop}
v^\phi_1(r, z) = - \frac{ A_\phi}{2r v^\phi_0} \partial_r v^\phi_0 + 2 rh_0   \partial_r v^\phi_0.
\end{equation}
Note that the above formula involves both geometric and hydrodynamical factors.

Acceptable 1PN solutions should satisfy the following (quite stringent) conditions: \textit{i)} $1\gg{|U_0|}/{c^2}\gg{|U_1|}/{c^4}$, \textit{ii)} $2GM_{\textrm{c}}/c^2 \ll r_{\textrm{in}}$, \textit{iii)} $c \gg c_{\textrm{s}}$, where $c_{\textrm{s}}$ is a speed of sound. Sample numerical models that do satisfy the above conditions are presented in \cite{JMMP_1PN}. It turns out that the 1PN correction to the velocity can be of the order of 10\% of the Newtonian value. The reader interested in the details of these model may consult \cite{JMMP_1PN}. In the remainder of this paper we focus on the construction of virial tests that can be applied to our numerical scheme.

\section{Virial identities}

In this section we use Cartesian and cylindrical coordinates. It is implicitly assumed that Latin indices refer to Cartesian coordinates.

The virial identity that can be used to test the Newtonian solution (including the central point-mass) was obtained in \cite{mach_virial}. It reads
\begin{equation}
\label{newt_vir}
E_\mathrm{pot} +2E_\mathrm{kin} + 2E_\mathrm{therm} = 0,
\end{equation}
where $E_\mathrm{pot} = \frac{1}{2} \int_{\mathbb R^3} d^3 x \ \rho (U_0 - GM_\mathrm{c} / |\mathbf{x}|)$ is the total potential energy, $E_\mathrm{kin} = \frac{1}{2} \int_{\mathbb R^3} d^3 x \ \rho v_{0i}v^i_0$ is the bulk kinetic energy and $E_\mathrm{therm} = \frac{3}{2} \int_{\mathbb R^3} d^3 x \ p$ is the internal thermal energy. We assume, as a virial test parameter, the value
\[\epsilon = \left| \frac{E_\mathrm{pot} +2E_\mathrm{kin} + 2E_\mathrm{therm}}{E_\mathrm{pot}} \right|. \]

In order to obtain the post-Newtonian relations we rewrite Eqs.~(\ref{c}) and (\ref{d}) in a slightly different form. Equation Eqs.~(\ref{c}) can be written in Cartesian coordinates as
\begin{equation}
\nonumber
\Delta A^i = - 16 \pi G \rho_0 v_0^i.
\end{equation}
Equation (\ref{d}) is split in two parts:
\begin{eqnarray}
\Delta U_1^\prime & = & 4 \pi G \left( \rho_1 + 2p_0 + \rho_0 \left( h_0 - 2 U_0 + 2 r^2 (v_0^\phi)^2 \right) \right), \label{aaab} \\
\Delta U_1^{\prime \prime} & = & 4 \pi G M_\mathrm{c} U^\mathrm{D}_0(0) \delta (\mathbf x), \label{aaac}
\end{eqnarray}
where $U_1 = U_1^\prime + U_1^{\prime \prime}$. The solution for $U_1^{\prime \prime}$ is
\[ U_1^{\prime \prime} = - \frac{G M_\mathrm{c} U_0^\text{D}(\mathbf{0})}{|\mathbf x|}. \]

Consider a vector
\[ a^i = \left( x^l \partial_l A_k + \frac{1}{2} A_k \right) \partial^i A^k - \frac{1}{2} x^i \partial_l A^k \partial^l A_k. \]
Its divergence reads
\begin{equation}
\label{aaaf}
\partial_i a^i = \left( x^l \partial_l A_k + \frac{1}{2} A_k \right) \Delta A^k = - 16 \pi G \left( x^l \partial_l A_k + \frac{1}{2} A_k \right) \rho_0 v_0^k.
\end{equation}
For a finite disk ($\rho_0$ of compact support), $A_k$ tends to zero sufficiently fast, and
\[ |\mathbf x|^2 a^i \to 0, \quad \mathrm{as} \quad |\mathbf x| = \sqrt{x^2 + y^2 + z^2} \to \infty. \]
Thus, by integrating Eq.~(\ref{aaaf}) over $\mathbb R^3$, and making use of the Gauss theorem, we see that
\[ 0 = \int_{\mathbb R^3} d^3 x \left( x^l \partial_l A_k + \frac{1}{2} A_k \right) \rho_0 v_0^k. \] 
Integrating by parts, one can get rid of the term with $\partial_l A_k$. This yields
\[ 0 = \int_{\mathbb R^3} d^3 x \left( -\frac{5}{2}\rho_0 A_k v_0^k - x^l \partial_l \left( \rho_0 v_0^k \right) A_k \right), \]
where only $0^{th}$ order terms are differentiated.

The above relation can be also written in terms of the vector components in cylindrical coordinates. Because of symmetry assumptions, we have $A_k v_0^k = A_\phi v_0^\phi$, and
\[ x^l \partial_l \left( \rho_0 v_0^k \right) A_k = x^l \nabla_l \left( \rho_0 v_0^k \right) A_k  = \rho_0 A_\phi v_0^\phi + x^l \partial_l \left( \rho_0 v_0^\phi \right) A_\phi. \]
This yields
\[ 0 = \int_{\mathbb R^3} d^3 x \left( -\frac{7}{2} \rho_0 A_\phi v_0^\phi - x^l \partial_l \left( \rho_0 v_0^\phi \right) A_\phi \right). \]

The virial relation following from Eq.~(\ref{aaab}) can be obtained in a similar way. It is enough to consider the divergence of the vector
\[ b^i = \left( x^l \partial_l U_1^\prime + \frac{1}{2} U_1^\prime \right) \partial^i U_1^\prime - \frac{1}{2} x^i \partial_l U_1^\prime \partial^l U_1^\prime. \]
It yields
\[ \partial_i b^i = \left( x^l \partial_l U_1^\prime + \frac{1}{2} U_1^\prime \right) \Delta U_1^\prime, \]
and analogously
\[ 0 = \int_{\mathbb R^3} d^3 x \left( x^l \partial_l U_1^\prime + \frac{1}{2} U_1^\prime \right) \left( \rho_1 + 2p_0 + \rho_0 \left( h_0 - 2 U_0 + 2 r^2 (v_0^\phi)^2 \right) \right).  \]

Many different forms of the above relation can be obtained by `playing' with Eq.~(\ref{e}). A helpful relation that can be used here is
\[ -x^l \partial_l U_1^{\prime \prime} = U_1^{\prime \prime}. \]

In the analogy to the Newtonian case we choose as virial test parameters $\epsilon^\prime = |(\epsilon^{\prime}_a + \epsilon^{\prime}_b) / \epsilon^{\prime}_a | $ and $\epsilon^{\prime\prime} = |(\epsilon^{\prime\prime}_a +  \epsilon^{\prime\prime}_b) / \epsilon^{\prime\prime}_a| $, where
\begin{eqnarray}
\epsilon ^{\prime}_{a} &=& \int_{\mathbb R^3} d^3 x \ \frac{7}{2}\rho_0 A_{\phi} v_0^{\phi}, \\
\epsilon ^{\prime}_{b} &=& - \int_{\mathbb R^3} d^3 x \ x^l \partial_l \left( \rho_0 v_0^{\phi} \right) A_{\phi}, \\
\epsilon ^{\prime \prime}_{a} &=& - \int_{\mathbb R^3} d^3 x \ \frac{1}{2} U_1^\prime \left( \rho_1 + 2p_0 + \rho_0 \left( h_0 - 2 U_0 + 2 r^2 (v_0^\phi)^2 \right) \right), \\
\epsilon ^{\prime \prime}_{b} &=& \int_{\mathbb R^3} d^3 x \ x^l \partial_l U_1^\prime \left( \rho_1 + 2p_0 + \rho_0 \left( h_0 - 2 U_0 + 2 r^2 (v_0^\phi)^2 \right) \right).
\end{eqnarray}

In Table \ref{tabone} we report results of the convergence tests for a sample system. In our implementation, numerical precision is controlled by the resolution of the grid, the maximum number $L$ of Legendre polynomials used in the angular expansion of the solutions of the scalar and vector Poisson equations, and a value of the maximal difference between density distributions obtained in the last two consecutive iterations $ \rho_\mathrm{tol}$. (In each iteration we compute the quantity $ \rho_\mathrm{err} = \max_{i,j} | \rho^{(k+1)}_{i,j} - \rho^{(k)}_{i,j}|$. Here index $k$ numbers subsequent iterations; indices $i$ and $j$ refer to different grid nodes. The iteration procedure is stopped, when $\rho_\mathrm{err} \le \rho_\mathrm{tol}$.)

The results of the virial test do depend on the grid resolution and almost do not depend on the numbers of Legendre polynomials and the precision in convergence procedure.

\begin{table}[t]
\caption{\label{tabone}Typical dependence of the results on the resolution of the numerical grid, the maximum number of the Legendre polynomials $L$, and the tolerance coefficient $ \rho_\mathrm{tol}$. These results are obtained for a polytropic fluid with polytropic index $\gamma = 5/3$ and the Keplerian rotation law $v_0^\phi = \omega_0 / r^{3/2}$. The mass of the disk is $M_\mathrm{d} = 0.327 M_\mathrm{c}$, the inner and outer radii are $r_\mathrm{in} = 50 R_\mathrm{S}$ and $r_\mathrm{out} = 500 R_\mathrm{S}$ respectively.
}
\begin{center}
\lineup
\begin{tabular}{*{7}{l}}
\br  
Resolution & $L$ & $ \rho_\mathrm{tol}$& $\epsilon$ & $\epsilon^\prime$ & $\epsilon^{\prime\prime}$ \\
\hline
100   $\times$ 100   & 100 & $10^{-6}$ & $2.53 \times 10^{-5}$ & $2.29 \times 10^{-4}$ & $3.51 \times 10^{-4}$\\
200   $\times$ 200   & 100 & $10^{-6}$ & $6.36 \times 10^{-6}$ & $5.87 \times 10^{-5}$ & $8.87 \times 10^{-5}$\\
400   $\times$ 400   & 100 & $10^{-6}$ & $1.56 \times 10^{-6}$ & $1.93 \times 10^{-5}$ & $2.29 \times 10^{-5}$\\
800   $\times$ 800   & 100 & $10^{-6}$ & $3.66 \times 10^{-7}$ & $3.65 \times 10^{-6}$ & $4.62 \times 10^{-6}$\\
1200 $\times$ 1200 & 100 & $10^{-6}$ & $1.44 \times 10^{-7}$ & $1.56 \times 10^{-6}$ & $2.49 \times 10^{-6}$\\
1600 $\times$ 1600 & 100 & $10^{-6}$ & $6.66 \times 10^{-8}$ & $5.43 \times 10^{-7}$ & $1.39 \times 10^{-6}$\\
\mr
400  $\times$ 400  & 50   & $10^{-6}$ & $1.56 \times 10^{-6}$ & $1.93 \times 10^{-5}$ & $2.27 \times 10^{-5}$\\
400  $\times$ 400  & 75   & $10^{-6}$ & $1.56 \times 10^{-6}$ & $1.93 \times 10^{-5}$ & $2.29 \times 10^{-5}$\\
400  $\times$ 400  & 100 & $10^{-6}$ & $1.56 \times 10^{-6}$ & $1.93  \times 10^{-5}$ & $2.29 \times 10^{-5}$\\
400  $\times$ 400  & 125 & $10^{-6}$ & $1.56 \times 10^{-6}$ & $1.93  \times 10^{-5}$ & $2.30 \times 10^{-5}$\\
400  $\times$ 400  & 150 & $10^{-6}$ & $1.56 \times 10^{-6}$ & $1.93  \times 10^{-5}$ & $2.30 \times 10^{-5}$\\
\mr
400  $\times$ 400  & 100 & $10^{-5}$ & $1.52 \times 10^{-6}$ & $1.94 \times 10^{-5}$ & $2.29 \times 10^{-5}$\\
400  $\times$ 400  & 100 & $10^{-6}$ & $1.56 \times 10^{-6}$ & $1.93 \times 10^{-5}$ & $2.29 \times 10^{-5}$\\
400  $\times$ 400  & 100 & $10^{-7}$ & $1.59 \times 10^{-6}$ & $1.93 \times 10^{-5}$ & $2.29 \times 10^{-5}$\\
400  $\times$ 400  & 100 & $10^{-8}$ & $1.59 \times 10^{-6}$ & $1.93 \times 10^{-5}$ & $2.29 \times 10^{-5}$\\
\br 
\end{tabular}
\end{center}
\end{table}

\ack
This research was carried out with the supercomputer `Deszno' purchased thanks to the financial support of the European Regional Development Fund in the framework of the Polish Innovation Economy Operational Program (contract no.\ POIG.\ 02.01.00-12-023/08). The work of PJ was partially supported by the Polish NCN grant \textit{Networking and R\&D for the Einstein Telescope}. PM and MP acknowledge the support of the Polish Ministry of Science and Higher Education grant IP2012~000172 (Iuventus Plus).

\end{document}